\documentclass[sigconf]{acmart}

\usepackage{natbib} 
\usepackage{censor}
\usepackage[utf8]{inputenc} 
\usepackage[T1]{fontenc}    
\usepackage{hyperref}       
\usepackage{url}            
\usepackage{booktabs}       

\usepackage{nicefrac}       
\usepackage{microtype}      
\usepackage{xcolor}         
\usepackage{amsmath,amsfonts}
\usepackage{graphicx}
\usepackage{blindtext}
\usepackage{textcomp}
\usepackage[inkscapelatex=false]{svg}
\usepackage{verbatim}
\usepackage{color,soul}
\usepackage{graphics}
\usepackage{listings}
\usepackage{multirow}
\usepackage{tabularx}

\usepackage[toc,page]{appendix}
\usepackage{mwe}%
\usepackage{todonotes}
\usepackage{tikz}
\usetikzlibrary{arrows.meta, positioning, shapes.multipart, fit, calc}
\usepackage{algorithm}
\usepackage{algpseudocode}
\usepackage{pgfplots}
\usepackage{siunitx}
\pgfplotsset{compat=newest}
\usepgfplotslibrary{groupplots}
\definecolor{darkred}{rgb}{0.55,0.0,0.0}
\pgfplotsset{compat=1.18}  

\usepackage{url}
\usepackage{hyperref}
\usepackage{breakurl} 
\usepackage{xurl}

\def\BibTeX{{\rm B\kern-.05em{\sc i\kern-.025em b}\kern-.08em
    T\kern-.1667em\lower.7ex\hbox{E}\kern-.125emX}}

\usepackage{xcolor}
\usepackage{xspace}

\newif\ifdraft
\drafttrue

\ifdraft
  \newcommand{\jhanote}[1]{{\textcolor{red}{ ***Shantenu: #1 }}\xspace}
\else
 \newcommand{\jhanote}[1]{}
 \fi

\def\BibTeX{{\rm B\kern-.05em{\sc i\kern-.025em b}\kern-.08em
    T\kern-.1667em\lower.7ex\hbox{E}\kern-.125emX}}

\AtBeginDocument{%
  \providecommand\BibTeX{{%
    Bib\TeX}}}

\setcopyright{acmlicensed}
\copyrightyear{2026}
\acmYear{2026}
\acmDOI{XXXXXXX.XXXXXXX}
\acmConference[SOCC]{ACM Symposium on Cloud Computing 2026}{November 18--20, 2026}{Singapore}
\acmISBN{978-1-4503-XXXX-X/2026/11}

\begin{document}

\title{OpRAG: A Resource-Deterministic Runtime for GPU-Backed Multi-Stage RAG Workflows}


\author{Arup Kumar Sarker}
\correspondingauthor
\email{djy8hg@virginia.edu}
\orcid{1234-5678-9012}
\author{Mills Staylor}
\email{qad5gv@virginia.edu}
\affiliation{%
  \institution{Biocomplexity Institute and Initiative}
  \institution{University of Virginia}
  \city{Charlottesville}
  \state{VA}
  \country{USA}
}

\author{Gregor von Laszewski}
\email{laszewski@gmail.com}
\author{Geoffrey Fox}
\email{vxj6mb@virginia.edu}
\affiliation{%
  \institution{Biocomplexity Institute and Initiative}
  \institution{University of Virginia}
  \city{Charlottesville}
  \state{VA}
  \country{USA}
}

\author{Aymen Alsaadi}
\email{aymen.alsaadi@rutgers.edu}
\author{Shantenu Jha}
\email{shantenu.jha@rutgers.edu}

\affiliation{%
  \institution{Rutgers University}
  \institution{Princeton Plasma Physics Laboratory}
  \city{Princeton}
  \state{NJ}
  \country{USA}
}

\begin{abstract}
Agentic retrieval-augmented generation (RAG) systems combine preprocessing, embedding, retrieval, memory access, context construction, generation, and vector-index updates. Although LLM decoding is GPU-bound, the surrounding orchestration layer can still limit end-to-end performance through serialization overhead, fragmented scheduling, inefficient batching, and CPU--GPU pipeline stalls. Existing frameworks provide flexible control flow, while distributed runtimes provide scalable task parallelism, but neither exposes RAG stages as resource-aware operators with deterministic execution semantics. We present OpRAG, a resource-deterministic distributed runtime for GPU-backed multi-stage RAG workflows. OpRAG models embedding, retrieval, reasoning, memory, and upsert as first-class operators and lowers them into communication-aware execution graphs. It combines an Arrow zero-copy data plane, persistent workers, bounded queues, CPU tokenizer prefetching, batched GPU embedding, and overlapped retrieval/generation execution to reduce non-model overhead around LLM inference. We evaluate OpRAG using Llama3-8B and Mistral-7B with FlashAttention~2, BF16 execution, and 32K RAG chunks. In end-to-end GPU pipeline experiments, OpRAG improves over the nearest competitor by 16.16\% for Llama3-8B and 15.66\% for Mistral-7B, and over RayScalableRAG by 20.57\% and 20.71\%, respectively. Against LangChain, LangGraph, CrewAI, and AutoGen, OpRAG is 17.77\% and 17.48\% faster than the best framework baseline. In Higress-style query serving, OpRAG reduces hybrid retrieval latency by 59.20--59.62\% and generation-scenario latency by 52.48--53.55\%, while preserving 100\% Recall@5. These results show that optimizing the distributed orchestration layer can substantially improve GPU-backed multi-stage RAG without modifying the LLM decoding kernel.
\end{abstract}

\ccsdesc[500]{Computer systems organization~Distributed architectures}
\ccsdesc[300]{Information systems~Data management systems}
\ccsdesc[300]{Information systems~Information retrieval}
\ccsdesc[100]{Computing methodologies~Parallel computing methodologies}

\maketitle

\section{Introduction}

Large language models (LLMs) are increasingly deployed within \emph{agentic workflows}, where retrieval, reasoning, tool invocation, and memory are dynamically orchestrated to solve knowledge-intensive tasks~\cite{arxiv2508}. Retrieval-Augmented Generation (RAG) extends LLMs with external knowledge by retrieving relevant context before generation~\cite{lewis2020retrieval,ibm25, intelvector}. In modern deployments, however, RAG is no longer a single retrieval call followed by decoding. A request may trigger document preprocessing, chunking, embedding, vector search, memory lookup, context construction, generation, and post-response index updates. These stages make agentic RAG a distributed systems problem, not only a model-serving problem.

This shift exposes two limitations in current systems. \textbf{First, data orchestration becomes a major bottleneck.} RAG pipelines move data across preprocessing, embedding, indexing, and retrieval stages, often through framework-specific objects, task queues, or external services. Distributed data systems such as Dask and
Spark provide scalable execution for static dataflow workloads, but they can incur scheduler, serialization, and object-management overhead in communication-bound regimes~\cite{arxiv2406,liang2025dataflow, abeykoon2022high}. \textbf{Second, dynamic workflow control limits reproducibility and optimization.} Agent frameworks such as LangChain and LangGraph provide flexible control flow, but LLM-driven decisions often determine execution routes at runtime~\cite{arxiv2508}. This makes RAG pipelines difficult to profile, replay, and optimize under fixed
resource budgets~\cite{arxiv2509,llamaindex25}.

Existing systems address different parts of this stack. Agent frameworks provide programming abstractions for tool use and multi-step reasoning, but they do not expose physical execution plans for embedding, retrieval, memory, and upsert stages. Distributed runtimes such as Ray and Dask provide task parallelism, but their execution models are not specialized for agentic RAG operators~\cite{moritz2018ray}. Retrieval gateways and vector databases optimize query-time lookup, but they usually treat preprocessing, memory management, and index updates as separate services~\cite{llamaindex25,rishabh25}. LLM serving systems such as vLLM and SGLang optimize decoding, batching, and KV-cache management, which is complementary to the orchestration problem studied
here~\cite{vllm,sglang}. As a result, GPU-backed RAG still lacks a unified execution model for coordinating CPU-side data movement, GPU embedding/generation, retrieval, memory, and vector-index updates.

We introduce \textbf{OpRAG}, a resource-deterministic distributed runtime for GPU-backed multi-stage RAG workflows. OpRAG treats the major stages of RAG as first-class operators:
\[
Op_{embed},\; Op_{retrieve},\; Op_{reason},\; Op_{memory},\;
Op_{upsert}.
\]
Each operator has explicit input/output schemas, resource requirements, and communication behavior. Following the observation that agentic operations can be mapped to broadcast, shuffle-compute, reduction, and
embarrassingly parallel execution patterns~\cite{arxiv2105}, OpRAG lowers workflow segments into communication-aware execution graphs. The runtime combines an Arrow/Cylon zero-copy data plane, persistent workers, bounded queues, CPU tokenizer prefetching, batched GPU embedding, and overlapped retrieval/generation execution to reduce
non-model overhead around LLM inference.

A key idea in OpRAG is to separate \emph{logical} agent decisions from \emph{physical} resource scheduling. OpRAG does not compile an entire agent conversation into one static graph. Instead, it compiles deterministic execution segments. 
OpRAG optimizes the non-model dataflow around decoding rather than modifying the LLM decoding kernel. It overlaps CPU-side preparation with GPU execution, uses batched embedding to reduce launch and Python
overhead, applies bounded queues for backpressure, and overlaps retrieval/generation execution to reduce idle time between CPU and GPU stages.

We evaluate OpRAG on A100 GPUs using Llama3-8B~\cite{dubey2024llama3} and Mistral-7B~\cite{jiang2023mistral} with FlashAttention~2 and BF16 execution~\cite{dao2023flashattention2}. In end-to-end GPU pipeline experiments over 32K RAG chunks, OpRAG improves over the nearest competitor by 16.16\% for Llama3-8B and 15.66\% for Mistral-7B, and over RayScalableRAG by 20.57\% and 20.71\%, respectively. In framework-level GPU comparisons, OpRAG is 17.77\% faster than the best Llama3-8B baseline and 17.48\% faster than the best Mistral-7B baseline. In Higress-style query serving, OpRAG reduces hybrid retrieval latency by 59.20--59.62\% and generation-scenario latency by 52.48--53.55\%, while preserving 100\% Recall@5 relative to Higress hybrid retrieval.

This paper makes the following contributions:

\begin{enumerate}
    \item \textbf{Multi-stage RAG operator runtime.}
    We formalize embedding, retrieval, reasoning, memory, and upsert as
    first-class operators and show how dynamic multi-stage RAG workflows can
    be executed through repeated deterministic DAG segments.

    \item \textbf{Resource-deterministic CPU/GPU scheduling.}
    We design a runtime that separates logical agent control from
    physical scheduling and uses persistent workers, bounded queues,
    tokenizer prefetching, GPU batching, and CPU--GPU overlap to reduce
    orchestration overhead.

    \item \textbf{Stateful retrieval and memory-aware execution.}
    We integrate retrieval, memory lookup, context construction, and
    index updates into one execution model, exposing both latency gains
    and retrieval-quality tradeoffs.

    \item \textbf{End-to-end GPU evaluation.}
    We evaluate OpRAG with Llama3-8B and Mistral-7B on A100 GPUs against
    agent frameworks, distributed RAG baselines, and Higress-style query
    serving, showing that optimizing the orchestration layer improves
    GPU-backed multi-stage RAG without modifying the LLM decoding kernel.
\end{enumerate}

\section{Background and Motivation}
\label{sec:background}

Agentic retrieval-augmented generation (RAG) systems combine large
language models with external knowledge, tool use, memory, and iterative
control flow. Classical RAG retrieves relevant documents before generation
~\cite{lewis2020retrieval}, while modern agentic
systems repeatedly interleave retrieval, reasoning, memory lookup, and
state updates~\cite{arxiv2508,yao2023react,shinn2023reflexion}. This
changes RAG from a single model-serving request into a distributed
execution problem: one query may involve preprocessing, chunking,
embedding, vector search, context construction, generation, memory
updates, and vector-index upserts.

\subsection{GPU-Backed multi-stage RAG}

LLM serving systems such as vLLM and SGLang improve decoding throughput
through batching, scheduling, and KV-cache management~\cite{vllm,sglang}.
These optimizations are essential, but they target the model-serving
kernel rather than the data-orchestration path around the model. In
multi-stage RAG, the end-to-end latency of a request can be expressed in equation~\ref{eq:trag}.
Even when \(T_{generate}\) is GPU-bound, the remaining stages can stall
the GPU or reduce throughput. Tokenization, embedding preparation,
retrieval over vector shards, memory lookup, context construction, and
index updates often execute on CPUs or external services~\cite{sarker2026aaflow}. If these stages
are scheduled independently, GPUs may wait for input batches while CPU workers wait for generation or vector-store commits.

This issue becomes more visible with production-scale models such as Llama3-8B and Mistral 7B~\cite{dubey2024llama3,jiang2023mistral}. Modern
GPU execution techniques such as FlashAttention~2 and BF16 execution improve model throughput~\cite{dao2023flashattention2}, but they do not eliminate serialization, scheduler overhead, or CPU--GPU pipeline stalls in the retrieval and embedding path. Therefore, GPU-backed multi-stage RAG requires a runtime that coordinates CPU-side data movement and GPU-side model execution as one pipeline.

\subsection{Limitations of Existing Execution Models}

Existing systems address different parts of the stack, but none provides a unified execution model for GPU-backed multi-stage RAG.

\textbf{Agent frameworks.}
Frameworks such as LangChain, LangGraph, CrewAI, and AutoGen provide flexible abstractions for tool use, multi-agent coordination, and iterative reasoning~\cite{langchain2023,langgraph2024,duan2024exploration,
crewai_kickoff_async,wu2024autogen}. However, they primarily operate at the application layer. Physical execution of embedding, retrieval, memory access, and upsert is delegated to external libraries or services, giving the runtime limited control over batching, data movement, worker placement, and resource-deterministic scheduling.

\textbf{Distributed data and task runtimes.}
Systems such as Spark, Flink, Dask, Ray, and Radical/Deep-Cylon provide scalable execution for data-parallel or task-parallel workloads~\cite{zaharia2016apache,
carbone2015apache,rocklin2015dask,moritz2018ray, sarker2024radical, osti_10634837}. These systems are
powerful substrates, but their abstractions are not specialized for the operator structure of multi-stage RAG. Retrieval, memory, reasoning-context construction, and index update are not first-class execution operators with explicit resource domains and communication semantics.

\textbf{Vector retrieval and gateway systems.}
Vector databases and retrieval gateways optimize similarity search, routing, and caching~\cite{johnson2019faiss,pinecone2023,lin2025higress,
higress_repo,higress_cache}. These systems improve query-time retrieval, but they typically treat ingestion, embedding, memory management, and index updates as separate components. As a result, they do not provide a single runtime abstraction for coordinating the full path from document preprocessing to retrieval, reasoning, memory, generation, and upsert.

\textbf{LLM serving runtimes.}
LLM serving systems optimize token generation using model-level
scheduling and memory management~\cite{vllm,sglang}. OpRAG is
complementary to these systems. Rather than replacing the LLM serving engine, OpRAG targets the distributed dataflow around it: embedding batches, retrieval paths, memory lookup, context construction, and index updates.

These limitations motivate OpRAG's design. Multi-stage RAG needs a runtime that treats embedding, retrieval, reasoning, memory, and upsert as systems operators rather than ad hoc callbacks. Each operator has a natural execution pattern: embedding is embarrassingly parallel and benefits from GPU batching; retrieval requires partition-aware routing and top-\(k\) reduction; reasoning performs bounded context aggregation; memory lookup and update require state-aware retrieval; and upsert requires batched writes to partitioned vector indices. OpRAG adapts
high-performance data movement ideas from Arrow/Cylon and
communication-aware dataframe systems~\cite{widanage2020high,
abeykoon2022high,shan2022hybrid,alsaadi2025rhapsody} to this
GPU-backed multi-stage RAG setting.

\section{Design}
\label{sec:design}

OpRAG is a resource-deterministic runtime for GPU-backed multi-stage
RAG workflows. Its design follows three principles: (1) represent
embedding, retrieval, reasoning, memory, and upsert as explicit
operators rather than framework callbacks; (2) separate dynamic agent
planning from physical resource scheduling; and (3) coordinate
CPU-side preprocessing, retrieval, memory access, and vector-index
updates with GPU-backed embedding and generation.

\begin{figure}[htpb]
    \begin{center}
    \includegraphics[width=1.0\linewidth]{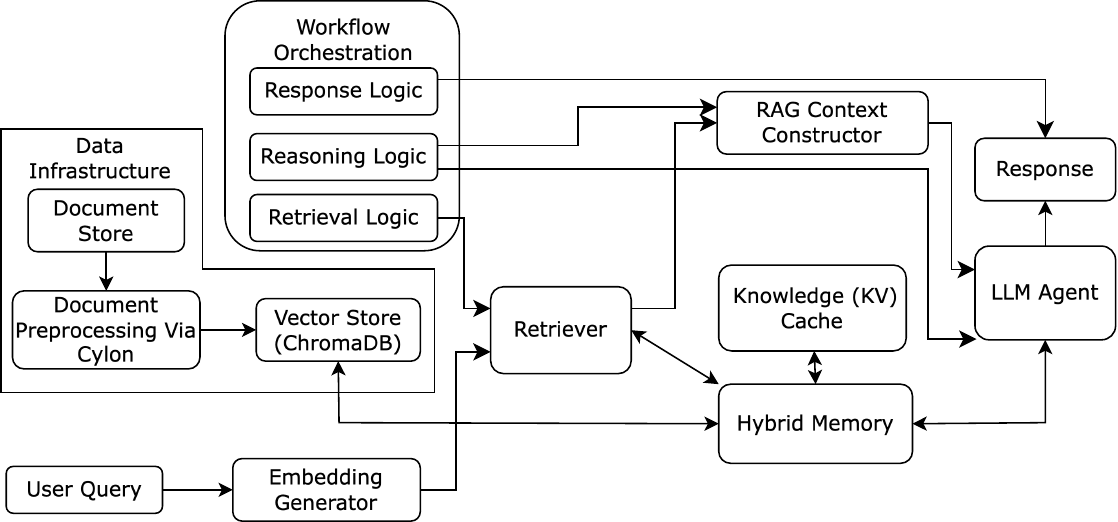}
    \caption{End-to-end OpRAG component flow with memory-aware RAG execution.
    The retriever accesses both static knowledge and memory state; the RAG
    context constructor combines them before generation, and reusable state can
    be written back through the memory/upsert path. } 
    \label{fig:oprag_component_flow}
    \end{center}
\end{figure}

\subsection{Overview}
Conventional RAG systems are often implemented as loosely coupled
pipelines that connect preprocessing, embedding, retrieval, generation,
and indexing through framework-specific objects or external services
~\cite{lewis2020retrieval,llamaindex25}. This
works at small scale, but creates fragmented data movement, weak
resource control, and non-deterministic execution in GPU-backed multi-stage
settings~\cite{arxiv2406,rishabh25,arxiv2508,liang2025dataflow}.
OpRAG addresses these limitations by lowering agentic RAG stages into
typed operator graphs over explicit resource domains.

Figure~\ref{fig:oprag_component_flow} shows the end-to-end component
flow. A user query is first embedded and routed to the retriever. The
retriever accesses both the static knowledge index and the memory module,
which stores prior interaction state, reusable summaries, and
intermediate reasoning artifacts. The RAG context constructor then merges
retrieved knowledge and memory state into a bounded context for the LLM
agent. After generation, reusable state can be promoted back into memory
and inserted through the upsert path. This figure is important because it
shows why memory is not an external cache in OpRAG: memory is accessed
through the same retrieval path and is scheduled as part of the runtime. At the runtime level, each visible component in
Fig.~\ref{fig:oprag_component_flow} maps to an OpRAG operator:
embedding maps to \(Op_{embed}\), retrieval maps to
\(Op_{retrieve}\), context construction maps to \(Op_{reason}\),
memory lookup and update map to \(Op_{memory}\), and vector-index
writes map to \(Op_{upsert}\). 

\begin{figure}[t]
    \centering
    \includegraphics[width=0.95\linewidth]{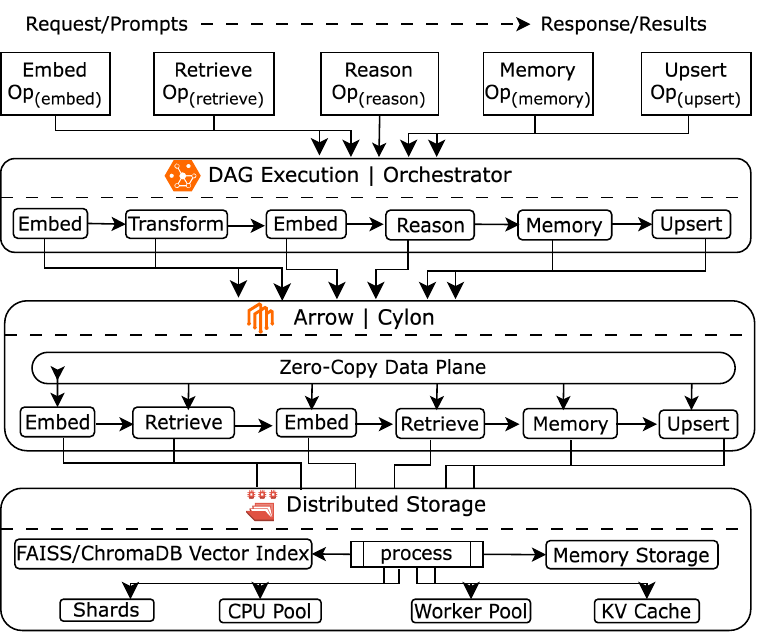}
    \caption{Layered OpRAG design. Dynamic Multi-Stage decisions are converted into deterministic operator segments. Operators execute over a shared Arrow/Cylon data plane, GPU-backed embedding and generation workers,
    partitioned vector indices, and a stateful memory layer.}
    \label{fig:oprag_layered_design}
\end{figure}

Figure~\ref{fig:oprag_layered_design} shows the layered design. The
agentic layer decides \emph{what} action is needed: retrieval, memory
lookup, query reformulation, reasoning, or state update. The OpRAG
runtime decides \emph{how} that action is executed by assigning
operators to CPU workers, GPU workers, distributed vector shards, or
memory stores. This separation preserves multi-stage flexibility while
exposing deterministic execution boundaries to the runtime.

\subsection{Multi-stage RAG Operator Model}
\label{subsec:operator_model}

OpRAG models a multi-stage RAG workflow as a set of first-class operators:
\[
\mathcal{W} =
\{Op_{embed}, Op_{retrieve}, Op_{reason}, Op_{memory}, Op_{upsert}\}.
\]
Each operator is defined as:
\[
Op_i = (I_i, O_i, f_i, P_i, R_i),
\]
where \(I_i\) and \(O_i\) are input and output objects, \(f_i\) is the operator function, \(P_i\) is the communication pattern, and \(R_i\) is the resource domain, such as CPU partitions, GPU workers, vector shards, or memory stores.

\textbf{Embedding.}
\(Op_{embed}\) maps text chunks to dense vectors:
\[
E = Op_{embed}(D; b_e),
\]
where \(D\) is a partitioned document/chunk collection and \(b_e\) is
the embedding batch size. Since chunks are independent, this operator is
embarrassingly parallel and benefits from CPU tokenizer prefetching and
batched GPU embedding~\cite{6972327}.

\textbf{Retrieval.}
\(Op_{retrieve}\) performs distributed top-\(k\) search over partitioned
vector indices:
\[
R_k = Op_{retrieve}(q,\mathcal{I}).
\]
The query is broadcast to index shards, each shard computes local
candidates, and OpRAG merges local results through a global top-\(k\)
reduction. This makes retrieval a partition-aware communication
operator rather than a black-box vector-store call.

\textbf{Reasoning.}
\(Op_{reason}\) converts retrieved evidence into a bounded context:
\[
C = Op_{reason}(R_k,q).
\]
The operator filters, ranks, and aggregates retrieved fragments before
passing a deterministic context payload to the LLM. This models context
construction as a reduction over evidence rather than a free-form
callback.

\textbf{Memory.}
\(Op_{memory}\) manages short-term state, intermediate artifacts, and
long-term vectorized summaries:
\[
(M',C_m)=Op_{memory}(M,C,q).
\]
Lookup follows the same distributed retrieval path as static knowledge
search, while updates are selectively promoted and later inserted through
the upsert path.

\textbf{Upsert.}
\(Op_{upsert}\) inserts or updates embeddings in partitioned vector
indices:
\[
\mathcal{I}' = Op_{upsert}(\mathcal{I},E;b_u),
\]
where \(b_u\) is the upsert batch size. Updates are routed to destination shards, coalesced, and committed in bulk to reduce write amplification and per-item index-update overhead~\cite{johnson2019faiss}.

\subsection{Segment-Level Compilation for Dynamic Multi-Stage Execution}
\label{subsec:segment_compilation}

\begin{figure}[t]
    \centering
    \includegraphics[width=0.99\linewidth]{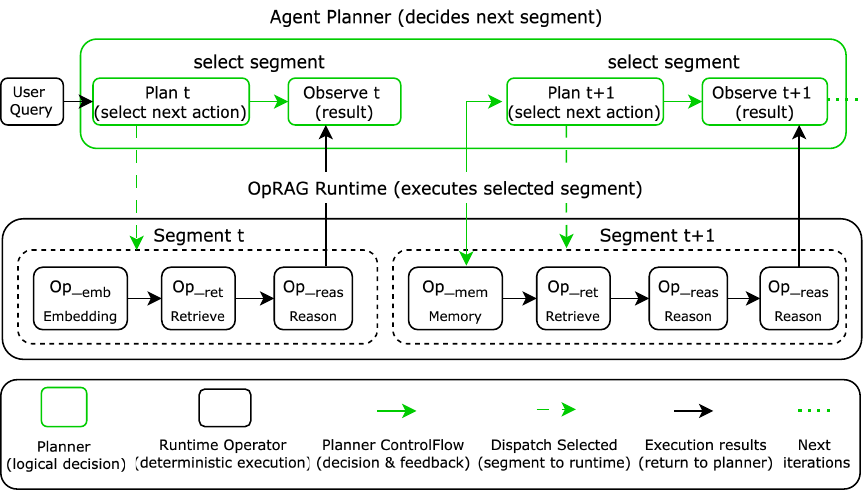}
    \caption{Dynamic agent execution in OpRAG. The Workflow Planner selects
    the next logical action after each observation, while the OpRAG Runtime
    executes the selected action as a deterministic operator segment.}
    \label{fig:dynamic_segments}
\end{figure}

A key challenge is that multi-stage workflows are dynamic, while distributed execution requires predictable runtime plans. OpRAG resolves this by compiling \emph{execution segments}, not entire conversations. At step \(t\), the planner observes state \(S_t\), emits a workflow segment
\(\mathcal{W}_t\), and OpRAG compiles it into:
\[
\mathcal{G}_t = Compile(\mathcal{W}_t,S_t),
\]
then executes:
\[
S_{t+1}=Execute(\mathcal{G}_t,S_t).
\]
The next agent decision may produce a different segment
\(\mathcal{W}_{t+1}\), which is compiled separately. Thus, the agent
remains dynamic across iterations, while each runtime segment has fixed
resource and communication semantics. Figure~\ref{fig:dynamic_segments} illustrates this execution model. The Workflow planner selects the next logical action after each observation, while the OpRAG Runtime executes the selected action as a deterministic operator segment. For example, one segment may execute \(Op_{embed}\rightarrow Op_{retrieve}\rightarrow Op_{reason}\), while a later segment may execute
\(Op_{memory}\rightarrow Op_{retrieve}\rightarrow Op_{reason}\rightarrow
Op_{upsert}\). This design supports ReAct-style plan--execute--observe
loops without forcing the entire conversation into one static DAG.

\subsection{Shared Data-Plane Abstraction}
\label{subsec:data_plane_design}

OpRAG requires a shared data-plane abstraction because multi-stage RAG
operators exchange heterogeneous objects: raw text chunks, tokenized
inputs, embeddings, retrieval candidates, memory records, context
payloads, and index-update batches. If each stage materializes these
objects through framework-specific containers or Python object stores,
the runtime loses control over serialization, routing, and batching.
Therefore, OpRAG represents intermediate RAG state as typed,
partitioned records that can be routed across operators with explicit
communication semantics.

The data plane exposes three design properties. First, it preserves
\emph{operator ownership}: each record carries routing metadata that
identifies the worker, vector shard, or memory partition responsible for
the next operation. Second, it supports \emph{communication-aware
execution}: retrieval uses query broadcast and top-\(k\) reduction,
reasoning uses evidence aggregation, and upsert uses shuffle followed by
batched writes. Third, it defines a clear \emph{zero-copy boundary}:
OpRAG avoids framework-level object serialization across preprocessing,
embedding staging, retrieval metadata movement, and index-update staging,
while treating the vector-index handoff as an implementation boundary.

\subsection{Resource-Deterministic CPU/GPU Scheduling}
\label{subsec:resource_deterministic_design}

OpRAG separates logical agent behavior from physical scheduling. The
agent decides whether retrieval, memory lookup, generation, or update is
needed; the runtime decides where and how the corresponding operators
execute. This enables reproducible traces, controlled batching, and
predictable resource use. For an operator stage processing \(N\) items in batches of size \(b\)
with up to \(P\) workers, the batch cost is:
\begin{equation}
T_{\text{batch}} = \alpha + \beta b,
\end{equation}
where \(\alpha\) is fixed per-batch overhead and \(\beta\) is per-item
work. With batching and parallel workers:
\begin{equation}
T_{\text{stage}} \approx
\frac{N}{bP}(\alpha + \beta b)
= \frac{N\alpha}{bP} + \frac{N\beta}{P}.
\end{equation}
Across frameworks, total time also includes runtime overhead:
\begin{equation}
T \approx
\frac{N\beta}{P} + \frac{N\alpha}{bP} + \Omega,
\label{eq:runtime_overtime}
\end{equation}
where \(\Omega\) captures scheduler overhead, serialization,
object-store traffic, synchronization, and idle time. OpRAG reduces
\(\Omega\) through zero-copy exchange, persistent workers, bounded
queues, and explicit communication primitives. For GPU-backed RAG, the end-to-end critical path is:
\begin{multline}
T_{RAG} =
T_{load} + T_{transform} + T_{embed} + T_{retrieve} \\
+ T_{context} + T_{generate} + T_{upsert}.
\label{eq:trag}
\end{multline}
OpRAG does not modify the LLM decoding kernel. Instead, it reduces and overlaps the non-model terms. In steady state, the effective pipelined cost approaches:
\begin{multline}
T_{\text{pipe}} \approx
\max(T_{cpu\_prep},T_{gpu\_embed},T_{retrieve}+T_{context}, \\
T_{generate},T_{upsert})+\Omega.
\label{eq:tpipe}
\end{multline}

This model explains why CPU/GPU overlap matters: if tokenization,
retrieval, and upsert are serialized around GPU execution, the pipeline
is additive; if they are overlapped through bounded queues and persistent
workers, the critical path approaches the maximum stage cost plus
residual overhead.

\subsection{Memory-Aware Retrieval and State Management}
\label{subsec:memory_design}

OpRAG treats memory as a scheduled operator rather than a passive cache.
The memory layer stores short-term agent state, intermediate reasoning
artifacts, and long-term vectorized summaries. Memory lookup follows the
same distributed retrieval path as knowledge search; memory update occurs
after generation through selective promotion and \(Op_{upsert}\). This
keeps state management measurable and prevents memory from becoming an
untracked side channel. Memory growth is proportional to embedding
dimension, number of retained summaries, metadata overhead, and the
promotion policy; we quantify this cost in the implementation discussion.

\section{Implementation}
\label{sec:implementation}

OpRAG implements the operator model from Section~\ref{sec:design} as a
resource-deterministic distributed runtime for GPU-backed multi-stage RAG.
Instead of exposing retrieval, embedding, memory, and index update as
loosely coupled framework callbacks, OpRAG lowers each selected agent
segment into a typed execution graph. The graph vertices are instantiated
operators and the edges are typed data dependencies. The runtime then
assigns each operator to a resource domain---CPU workers, GPU workers,
distributed vector shards, or memory stores---and executes the segment
using bounded queues, persistent workers, explicit communication, and
batch-aware scheduling. The implementation has five components: (1) an operator runtime, (2) a zero-copy data plane, (3) a GPU-backed embedding and generation path, (4) a memory-aware retrieval and upsert path, and (5) an asynchronous CPU/GPU execution engine. Together, these components
realize the design goals of operator-driven execution, deterministic
resource scheduling, and reduced non-model overhead around LLM
inference.

\subsection{Operator Runtime}
\label{subsec:operator_runtime}

At runtime, OpRAG materializes each abstract operator
\[
Op_{embed},\; Op_{retrieve},\; Op_{reason},\; Op_{memory},\;
Op_{upsert}
\]
as a concrete runtime primitive with explicit input/output schemas,
partition ownership, batching policy, and resource placement. For each
agent step \(t\), the planner emits a workflow segment
\(\mathcal{W}_t\), and OpRAG lowers it into:
\[
G_t = (V_t, E_t, R_t),
\]
where \(V_t\) is the set of operator instances, \(E_t\) is the set of
typed dependencies, and \(R_t\) maps each operator to a resource domain.
This mapping is important because logical agent decisions and physical
execution are separated: the agent decides \emph{what} action is needed,
while OpRAG decides \emph{how} that action is scheduled.

\textbf{Embedding.}
The embedding operator receives partitioned text chunks and produces
dense vectors. Each chunk is independent, so the operator is scheduled as
an embarrassingly parallel batched map. In the GPU path, CPU workers
perform tokenization and length grouping, while GPU workers execute
batched embedding using the same HuggingFace model family used for
generation. The embedding path uses hidden-state extraction followed by
mean pooling and vector normalization. This preserves a consistent
semantic workload across baselines while allowing OpRAG to optimize
batch formation and CPU/GPU overlap.

\textbf{Retrieval.}
The retrieval operator dispatches a query vector to partition-local vector
indices. Each shard computes local top-\(k\) candidates, and the runtime
merges local candidates into a global ranked result. This implements
query broadcast followed by local search and top-\(k\) reduction. In the
Higress-style query benchmark, OpRAG also supports a dense-first hybrid
policy: FAISS first selects a dense top-\(N\) candidate set, and lexical
reranking is applied only to those candidates. This reduces retrieval cost
while allowing retrieval quality to be quantified through Recall@\(k\) and
Jaccard overlap against full hybrid retrieval.

\textbf{Reasoning.}
The reasoning operator converts retrieved evidence into a bounded context
object for the downstream LLM. Rather than treating context construction
as a free-form orchestration callback, OpRAG models reasoning as a
deterministic reduction over ranked evidence. Local evidence can be
filtered, scored, and compacted before the final context is assembled.
The output is a typed context payload containing text, provenance, score,
and memory-scope metadata.

\textbf{Memory.}
The memory operator manages short-term state, intermediate reasoning
artifacts, and long-term vectorized summaries. Lookup uses the same
partitioned retrieval path as static knowledge search. Update uses batched
state insertion and optional summary promotion. This makes memory a
scheduled operator rather than an ad hoc side channel. The runtime stores
retrieved chunk identifiers, source metadata, prior-turn summaries, and
optional vector embeddings for reusable context.

\textbf{Upsert.}
The upsert operator inserts new embeddings into partitioned vector
indices. OpRAG groups updates by destination shard and commits them in
bulk to reduce per-item write overhead and write amplification. The
operator follows a shuffle-reduce pattern: updates are routed to the
correct shard, coalesced, and inserted through the vector backend.

\subsection{Zero-Copy Data Plane}
\label{subsec:data_plane_impl}

The implementation realizes the data-plane abstraction using Apache Arrow for in-memory layout and Cylon for distributed partitioned execution~\cite{widanage2020high,perera2023depth,shan2022hybrid, staylor2025combining, perera2023supercharging}. Each document partition is stored as an Arrow-backed table with fields for source text, chunk identifier, provenance, routing key, memory scope, and optional vector payload. During ingestion, the load and transform stages produce Arrow records; the embedding stage appends vector fields to the same logical records; and the retrieval and upsert stages consume these records without converting them into framework-specific Python objects. Cylon provides the communication layer used by OpRAG operators. Retrieval uses broadcast to distribute query vectors to index shards and reduction to merge local top-\(k\) candidates. Context construction uses gather or reduce operations to assemble ranked evidence into a bounded prompt payload. Upsert uses shuffle to route vectors to their destination shards and then commits them in coalesced batches. These operations are backed by communication substrates such as MPI, UCX, and Gloo~\cite{dalcin2005mpi4py,shamis2015ucx,gloo:online}. 

The zero-copy claim is scoped to the OpRAG data plane. OpRAG avoids Python-object serialization for preprocessing output, embedding staging, retrieval metadata, context records, and index-update batches. At the FAISS boundary, vectors must be presented as contiguous \texttt{float32} buffers. When Arrow buffers are already contiguous and type-compatible, the handoff avoids additional Python-level serialization; however, FAISS may still internalize vectors depending on index type. Thus, OpRAG provides serialization-free exchange up to the vector-index handoff, not a claim that all FAISS internals are zero-copy~\cite{johnson2019faiss}.

\subsection{GPU-Backed Embedding and Generation Path}
\label{subsec:gpu_path}

The GPU implementation uses local HuggingFace causal language models for both embedding and generation. The evaluated models are Llama3-8B and Mistral-7B~\cite{dubey2024llama3,jiang2023mistral}. The runtime uses BF16 execution and FlashAttention~2~\cite{dao2023flashattention2}. Each Slurm task owns one GPU and loads one complete model replica. Scaling is data parallel across corpus shards. OpRAG does not rely on tensor parallelism or pipeline parallelism. Note that while production systems typically use smaller encoder models for embedding, we deliberately use the generative LLM for hidden-state extraction. This stress-tests GPU memory management and CPU-GPU overlap on homogeneous hardware, mirroring emerging trends in unified architectures (e.g., GritLM~\cite{muennighoff2025generative}). This design isolates the effect of the orchestration layer around GPU-backed RAG rather than changing the underlying LLM serving kernel. The embedding stage treats a causal language model as a hidden-state feature extractor. Each input chunk is processed as:
\begin{multline}
\text{chunk} \rightarrow \text{tokenize} \rightarrow
\text{model forward hidden states} \rightarrow \\
\text{mean pool} \rightarrow \text{L2 normalize}.
\end{multline}
The resulting normalized vectors are inserted into FAISS using the same
vector-store semantics as the baselines. No pre-embedded corpus is used;
embedding remains part of the measured RAG pipeline. OpRAG uses a length-bucket compiled embedding graph with CUDA graph repeated shape strategy which targets the dominant cost in the GPU pipeline: repeated embedding forward passes over many text chunks. It combines three optimizations. \textbf{First}, OpRAG applies length-bucketed batching. Chunks are grouped by
similar token length before embedding. Transformer batches are normally padded to the longest sequence in the batch, so mixing short and long chunks wastes attention and memory bandwidth. Length bucketing reduces padding waste and increases the fraction of useful tokens processed per GPU forward pass. \textbf{Second}, OpRAG uses a compiled embedding path. The embedding forward pass has a repeated structure: tokenize a batch, run the model with
\texttt{use\_cache=False}, pool the final hidden states, and normalize the vectors. Compiling this path reduces Python dispatch overhead and improves steady-state execution for repeated embedding batches. \textbf{Third}, OpRAG makes the embedding path CUDA-graph friendly. CUDA graphs are most effective when operation shapes are stable across iterations.
Length bucketing and fixed embedding batch sizes reduce shape variation, making repeated model forward calls more graph-friendly. This reduces CPU-side launch overhead and stabilizes GPU execution.

The purpose is not to change the model or the embedding semantics. The same model, corpus, chunk count, maximum input length, FAISS insertion logic, and generation settings are used across the compared systems. OpRAG improves performance by changing how batches are scheduled and executed, not by reducing the workload. The generation path uses batched autoregressive decoding through the same HuggingFace model replica. OpRAG does not modify the decoding kernel. Instead, it improves the query path by scheduling CPU and GPU work concurrently. While the GPU generates answers for batch \(n\), the
CPU prepares retrieval results and prompt contexts for batch \(n+1\):
\[
\text{CPU retrieval/context}_{n+1}
\parallel
\text{GPU generation}_{n}.
\]
This overlap reduces idle time between retrieval and generation. As a result, OpRAG improves end-to-end query latency through orchestration, batching, and pipeline scheduling, while keeping the underlying model execution backend unchanged.

\subsection{Memory-Aware Retrieval and Upsert}
\label{subsec:impl_memory_retrieval}

OpRAG maintains two logical vector indices: a static knowledge index and a memory index. The knowledge index stores corpus chunks. The memory index stores reusable summaries, prior-turn context, and vectorized interaction state. Both indices follow the same partitioning and routing model, so retrieval, memory lookup, and memory update can be scheduled as reproducible runtime operators. For standard retrieval, OpRAG performs partition-local dense search followed by global top-\(k\) reduction. For Higress-style hybrid retrieval, OpRAG computes dense and lexical scores over the same full candidate space as HigressRAG and then applies the same weighted score fusion:
\[
score(d,q) =
w_d \cdot \widehat{s}_{dense}(d,q) +
w_l \cdot \widehat{s}_{lexical}(d,q),
\]
where \(w_d\) and \(w_l\) are dense and lexical weights. The final result
is:
\[
TopK(q) =
TopK_{d \in \mathcal{I}}
\left(
w_d \widehat{s}_{dense}(d,q) +
w_l \widehat{s}_{lexical}(d,q)
\right).
\]

Dense scores are computed with vectorized matrix operations over the embedding matrix. Lexical scores are computed through an inverted-postings BM25 representation, which visits only documents that contain query terms while preserving exact full-corpus BM25 semantics. Final ranking uses partial top-\(k\) selection rather than a full sort over all candidates. 
Memory state is selectively promoted after generation. Short-lived execution traces remain local, while summaries and reusable context are inserted into the memory index through \(Op_{upsert}\). During context-dependent follow-up queries, \(Op_{memory}\) retrieves prior-turn state and augments the query context before retrieval or generation. This memory path is optional: self-contained queries can bypass it, while underspecified follow-up queries can use it to recover missing context. The memory cost can be estimated from the embedding dimension. A \(d\)-dimensional float32 vector requires \(4d\) bytes before index and metadata overhead. For \(d=768\), one vector requires approximately 3KB of raw vector storage; 10K memory summaries require roughly 30MB before FAISS and metadata overhead. 

\subsection{Asynchronous CPU/GPU Execution Engine}
\label{subsec:async_engine}

The ingestion pipeline is organized into four explicit stages:
\begin{equation}
Load \rightarrow Transform \rightarrow Embed \rightarrow Upsert.
\label{eq:four-pipeline}
\end{equation}

Unlike synchronous pipelines that insert barriers between stages, OpRAG
connects stages through bounded queues and persistent worker pools. The
load stage reads rank-local corpus files and materializes Arrow/Cylon
tables. The transform stage performs splitting, normalization, metadata
alignment, and chunk construction. The embed stage batches chunks and
executes GPU embedding. The upsert stage coalesces embeddings into larger
write batches and inserts them into partitioned FAISS indices.
OpRAG reduces \(\Omega\) using zero-copy exchange, persistent workers, bounded queues, and explicit operator scheduling in equation~\ref{eq:runtime_overtime}. Bounded queues provide backpressure and prevent unbounded intermediate state, while persistent workers avoid repeated startup and model-loading overhead. For GPU-backed RAG, the critical path is defined in equation~\ref{eq:trag}.
OpRAG optimizes the non-model terms and overlaps them with GPU execution
where possible. In steady state, the effective pipeline cost is calculated in the equation ~\ref{eq:tpipe}.
This is the central implementation idea: instead of serializing CPU
preparation, GPU execution, retrieval, and upsert, OpRAG overlaps them
under deterministic operator scheduling.

\subsection{Distributed Execution and Output Aggregation}
\label{subsec:distributed_execution}

The distributed GPU benchmark uses one Slurm task per GPU. Each task
loads one model replica, processes one rank-local corpus shard, writes
rank-local measurements, and participates in rank-0 aggregation. The
implementation records per-stage timings for load, transform, embed,
upsert, and generation. It also records generated tokens, generation
throughput, peak GPU memory, setup time, and time-to-first-token when
available. Rank-local outputs are written as CSV/JSON files and merged by
rank 0 into experiment summaries.

This execution model keeps the comparison controlled. All configurations
use the same model, corpus, chunk count, generation parameters, embedding
semantics, and vector insertion semantics. Differences therefore arise
from the runtime path: batching policy, scheduler overhead, object
movement, CPU/GPU overlap, and index-update strategy.

\subsection{Implementation Scope}
\label{subsec:impl_scope}

OpRAG is not an LLM serving kernel and does not replace systems such as
vLLM or SGLang~\cite{vllm,sglang}. Those systems optimize model serving,
KV-cache management, and structured decoding. OpRAG
optimizes the dataflow that surrounds GPU-backed RAG, including
preprocessing, embedding, retrieval, memory lookup, context construction,
and upsert. The runtime can therefore be used with existing model engines
and vector backends while providing deterministic execution semantics for
the orchestration layer.

\section{Evaluation}
\label{sec:evaluation}

We evaluate OpRAG to answer five questions.

\textbf{Q1:} Does resource-deterministic orchestration improve
end-to-end GPU-backed RAG pipelines with production-scale models?
\textbf{Q2:} Does OpRAG improve over agent-framework pipelines such
as LangChain, LangGraph, CrewAI, and AutoGen?
\textbf{Q3:} How does OpRAG compare with distributed RAG baselines
such as RayScalableRAG, DaskScalableRAG, AsyncParallelOnly,
and HigressRAG?
\textbf{Q4:} What latency/quality tradeoff does OpRAG introduce in
Higress-style query serving?
\textbf{Q5:} Do the CPU scaling experiments support the proposed
operator-runtime model beyond the two-GPU setting?

\subsection{Experimental Setup}
\label{subsec:eval_setup}

\textbf{GPU platform.}
GPU experiments run on the UVA Rivanna GPU partition using two
NVIDIA A100 GPUs. Each Slurm task owns one GPU and loads one
complete HuggingFace model replica. The runs use data parallelism
across corpus shards rather than tensor parallelism or pipeline
parallelism. This isolates the orchestration layer around RAG rather
than evaluating a new model-parallel serving engine.

\textbf{Models.}
We evaluate two production-scale open models: Llama3-8B
(\texttt{meta-llama/Meta-Llama-3-8B-Instruct})~\cite{dubey2024llama3}
and Mistral-7B (\texttt{mistralai/Mistral-7B-Instruct-v0.3})
~\cite{jiang2023mistral}. Both runs use BF16 execution and
FlashAttention~2~\cite{dao2023flashattention2}. The validated
FlashAttention version is \texttt{2.8.3.post1}.

\textbf{Dataset.}
The GPU pipeline and framework experiments use a public
HuggingFace text corpus derived from WikiText-103. The final
configuration contains 32K chunks, split across two rank-local shards:
16K chunks per GPU, 64 files per GPU, and 900 characters per chunk.
The generation workload uses 64 sampled prompts per GPU, maximum
input length 128 tokens, and maximum generation length 32 tokens.
The Higress-style query benchmark uses 1024 total queries per
scenario, maximum input length 512 tokens, and maximum generation
length 32 tokens.

\textbf{Baselines.}
We compare against three classes of systems. First, we compare
against agent-framework pipelines: LangChain~\cite{langchain2023},
LangGraph~\cite{langgraph2024}, CrewAI~\cite{crewai_kickoff_async},
and AutoGen~\cite{wu2024autogen}. Second, we compare against
distributed pipeline baselines: RayScalableRAG, built on
Ray~\cite{moritz2018ray}; DaskScalableRAG, built on
Dask~\cite{rocklin2015dask}; AsyncParallelOnly; and HigressRAG,
which follows the Higress-style gateway/retrieval path
~\cite{lin2025higress,higress_repo,higress_cache}. Third, for
query-serving experiments, we compare OpRAG against full
Higress-style hybrid retrieval.

\textbf{Controlled comparison.}
All GPU baselines use the same model, corpus, chunk count,
generation settings, embedding semantics, and FAISS vector insertion
semantics~\cite{johnson2019faiss}. Model loading and OpRAG compile
setup are reported separately and excluded from steady-state totals.
This makes the measured differences attributable to batching,
scheduling, CPU/GPU overlap, retrieval routing, and index-update
strategy rather than to model or dataset changes.

\subsection{Metrics}
\label{subsec:eval_metrics}

We report stage-wise runtime for \textsc{Load}, \textsc{Transform},
\textsc{Embed}, \textsc{Upsert}, and \textsc{Generate}, as well as
end-to-end total time. For GPU framework experiments, we also report
chunks processed per second:
\[
\text{Chunks/s} = \frac{N_{\text{chunks}}}{T_{\text{total}}}.
\]

For latency comparisons, relative improvement is:
\[
\text{Improvement} =
\frac{T_{\text{baseline}} - T_{\text{OpRAG}}}
     {T_{\text{baseline}}} \times 100.
\]

For query-serving quality, OpRAG compares its top-\(k\) retrieved
chunk IDs with the full HigressRAG hybrid retrieval result:
\[
Recall@k =
\frac{|TopK_{\text{Higress}} \cap TopK_{\text{OpRAG}}|}
     {|TopK_{\text{Higress}}|},
\]

For conversational retrieval, we also report Top-1 accuracy,
Hit@\(k\), and Mean Reciprocal Rank (MRR):
\[
MRR = \frac{1}{N}\sum_{i=1}^{N}\frac{1}{r_i},
\]
where \(r_i\) is the rank of the first correct result.

\subsection{GPU Pipeline Benchmark}
\label{subsec:gpu_pipeline}

Table~\ref{tab:gpu_pipeline} reports end-to-end GPU RAG pipeline
performance. The dominant stage is embedding: non-OpRAG systems spend
approximately 152--153 seconds in embedding, while OpRAG reduces this
to 126.832 seconds on Llama3-8B and 128.287 seconds on Mistral-7B.
Generation remains nearly identical across systems, around 4.2 seconds,
which confirms that the improvement comes from the orchestration and
embedding path rather than from modifying the decoding kernel.

\begin{table}[t]
\centering
\caption{GPU end-to-end pipeline benchmark on two A100 GPUs and 32K chunks with Llama3-8B(Row 2-6) and Mistral-7B (Row 7-11). Times are seconds. Tran = Transform, Gen = Generate}
\label{tab:gpu_pipeline}
\small
\setlength{\tabcolsep}{4pt}
\begin{tabular}{lrrrrrr}
\toprule
Baselines & Load & Trans & Embed & Upsert & Gen & Total \\
\midrule
AsyncParallelOnly  & 0.032 & 0.016 & 152.127 & 0.096 & 4.215 & 156.462 \\
DaskScalableRAG    & 0.186 & 0.033 & 152.135 & 0.101 & 4.206 & 156.615 \\
RayScalableRAG & 8.098 & 0.521 & 152.215 & 0.096 & 4.227 & 164.995 \\
HigressRAG         & 0.033 & 0.022 & 151.979 & 0.098 & 4.224 & 156.309 \\
\textbf{OpRAG}              & 0.030 & 0.019 & \textbf{126.832} & 0.344 & 4.193 & \textbf{131.048} \\
\midrule
AsyncParallelOnly  & 0.035 & 0.016 & 152.900 & 0.099 & 4.212 & 157.270 \\
DaskScalableRAG    & 0.197 & 0.031 & 152.770 & 0.098 & 4.199 & 157.278 \\
RayScalableRAG & 9.320 & 0.290 & 153.238 & 0.102 & 4.194 & 167.138 \\
HigressRAG         & 0.118 & 0.022 & 152.776 & 0.104 & 4.207 & 157.137 \\
\textbf{OpRAG}     & 0.031 & 0.019 & 128.287 & 0.281 & 4.184 & \textbf{132.525} \\
\bottomrule
\end{tabular}
\end{table}

For Llama3-8B, the nearest non-OpRAG competitor is HigressRAG
at 156.309 seconds. \textbf{OpRAG completes in 131.048 seconds, improving
total time by 16.16\%}. Relative to RayScalableRAG, the improvement is 20.57\%. For Mistral-7B, HigressRAG is again the nearest competitor at
157.137 seconds. OpRAG completes in 132.525 seconds, improving total
time by 15.66\%. Relative to RayScalableRAG, OpRAG improves by 20.71\%.

These results support the central idea of OpRAG: the runtime improves
GPU-backed RAG by optimizing the distributed orchestration around the
model. The largest gain appears in embedding because OpRAG uses
length-bucketed GPU batches, a compiled embedding path, CPU tokenizer
prefetch, and persistent workers. The slightly higher upsert time is
small compared with the embedding savings, so the end-to-end critical
path still improves.

\subsection{GPU Framework Benchmark}
\label{subsec:gpu_framework}

Table~\ref{tab:gpu_framework} evaluates the same GPU-backed workload
against agent framework pipelines. This experiment isolates whether the
OpRAG runtime improves over flexible application-level orchestration
systems when all frameworks use the same corpus, models, and generation
settings.

\begin{table}[t]
\centering
\caption{GPU framework benchmark on two A100 GPUs and 32K chunks with Llama3-8B(Row 2-6) and Mistral-7B (Row 7-11). Times are seconds.}
\label{tab:gpu_framework}
\small
\setlength{\tabcolsep}{4pt}
\begin{tabular}{lrrrrr}
\toprule
Framework & Embed & Upsert & Generate & Total & Chunks/s \\
\midrule
LangChain & 152.123 & 0.097 & 4.263 & 156.477 & 204.50 \\
LangGraph & 151.983 & 0.099 & 4.243 & 156.320 & 204.71 \\
CrewAI    & 152.203 & 0.099 & 4.244 & 156.546 & 204.41 \\
AutoGen   & 152.077 & 0.098 & 4.239 & 156.419 & 204.58 \\
\textbf{OpRAG}     & 124.306 & 0.540 & 4.237 & \textbf{128.543} & \textbf{248.94} \\
\midrule
LangChain & 151.140 & 0.113 & 4.112 & 155.425 & 205.89 \\
LangGraph & 151.037 & 0.098 & 4.103 & 155.282 & 206.08 \\
CrewAI    & 151.373 & 0.099 & 4.104 & 155.614 & 205.64 \\
AutoGen   & 151.239 & 0.101 & 4.101 & 155.489 & 205.80 \\
\textbf{OpRAG}     & 123.997 & 0.329 & 4.097 & \textbf{128.136} & \textbf{249.73} \\
\bottomrule
\end{tabular}
\end{table}

For Llama3-8B, LangGraph is the fastest baseline at 156.320 seconds,
while \textbf{OpRAG completes in 128.543 seconds with a 17.77\% improvement}. For Mistral-7B, LangGraph is again the fastest baseline at 155.282 seconds, while OpRAG completes in 128.136 seconds with a 17.48\%
The framework benchmark strengthens the pipeline result. 
LangChain, LangGraph, CrewAI, and AutoGen provide flexible agent control, but their physical execution remains callback-oriented. OpRAG preserves the same semantic workload but changes the physical schedule: CPU tokenization is prefetched, embedding batches are length-grouped, and execution uses persistent workers and bounded queues. The gain therefore comes from resource-deterministic execution rather than from weakening the workload.

\subsection{Higress-Style GPU Query Serving}
\label{subsec:higress_gpu}

The previous two experiments evaluate ingestion and pipeline execution. We next evaluate query serving against HigressRAG, the closest retrieval baseline. HigressRAG performs full hybrid retrieval by combining dense
FAISS scores and lexical scores over the corpus. OpRAG uses dense-first candidate pruning followed by lexical reranking over the top 128 dense candidates. This is a different physical retrieval schedule, so we report
both latency and quality relative to full HigressRAG hybrid retrieval.

\begin{table}[t]
\centering
\caption{Higress-style GPU query serving. Latencies are milliseconds.}
\label{tab:higress_gpu}
\small
\begin{tabular}{llrrrr}
\toprule
Model & Scenario & Higress & OpRAG & Impr. & Recall@5 \\
\midrule
Llama3-8B & Hybrid retrieval & 69.87 & 28.21 & \textbf{59.62\%} & 1.000 \\
Llama3-8B & LLM generation   & 217.41 & 103.30 & \textbf{52.48\%} & 1.000 \\
Mistral-7B & Hybrid retrieval & 68.63 & 28.00 & \textbf{59.20\%} & 1.000 \\
Mistral-7B & LLM generation   & 216.47 & 100.55 & \textbf{53.55\%} & 1.000 \\
\bottomrule
\end{tabular}
\end{table}

For Llama3-8B, OpRAG reduces hybrid retrieval latency from 69.87 ms to 28.21 ms, and reduces the retrieval-plus-generation path from 217.41 ms to 103.30 ms. For Mistral-7B, OpRAG reduces hybrid retrieval from 68.63 ms to 28.00 ms, and reduces the generation scenario from 216.47 ms to 100.55 ms. Recall@5 remains 1.0, with an average overlap of 5/5 retrieved chunks.
The results show that OpRAG improves both retrieval-only and generation-heavy scenarios. Retrieval-only gains come from exact operator-level improvements: postings-list BM25, vectorized dense scoring, and partial top-k selection. Generation-heavy gains come from batching and CPU/GPU overlap. These optimizations are complementary: retrieval becomes faster, and the remaining retrieval/context work is hidden behind batched GPU generation when possible.

\subsection{CPU Scaling and Microarchitectural Results}
\label{subsec:cpu_results}

Because RAG orchestration bottlenecks are overwhelmingly in CPU-side data movement, the distributed scaling evaluation focuses on the CPU cluster, while the GPU setup serves purely as a microbenchmark for the CPU-GPU handover. We also retain CPU and multicore experiments as supporting evidence for the operator-runtime design. These experiments isolate pipeline scheduling, zero-copy movement, bounded queues, and upsert coalescing without GPU model execution.

\begin{table}[t]
\centering
\caption{CPU ingestion benchmark with 10M chunks and 4096 files. Times are
seconds.}
\label{tab:cpu_ingestion}
\small
\begin{tabular}{lrrrrr}
\toprule
Config & Load & Trans. & Embed & Upsert & Total \\
\midrule
RayScalableRAG & 38.411 & 2.003 & 3.105 & 4.509 & 48.854 \\
AsyncParallelOnly  & 2.646 & 2.047 & 8.067 & 0.724 & 11.641 \\
DaskScalableRAG    & 2.876 & 2.194 & 3.222 & 12.423 & 16.188 \\
HigressRAG         & 0.890 & 2.108 & 1.765 & 0.634 & 4.439 \\
\textbf{OpRAG}     & 0.952 & 1.277 & 1.507 & 0.437 & \textbf{3.487} \\
\bottomrule
\end{tabular}
\end{table}

Table~\ref{tab:cpu_ingestion} shows that \textbf{OpRAG improves the CPU pipeline over DaskScalableRAG by \(16.188/3.487=4.64\times\) and over HigressRAG by \(4.439/3.487=1.28\times\)}. The largest differences are in
Transform, Embed, and Upsert. This supports the same mechanism observed on GPUs: OpRAG is not merely adding parallelism; it changes the execution model by overlapping stages, coalescing writes, and reducing runtime
coordination overhead.

The CPU scaling experiments use physical workers on a 40-core-per-node cluster. In the 1024-worker strong-scaling experiment, OpRAG uses 26 nodes and 1024 physical workers. Pipeline latency decreases from 13.653 seconds at 256 workers to 4.505 seconds at 1024 workers, while remaining ahead of the nearest baseline. This result supports the claim that the operator runtime scales beyond a single GPU node and that the bounded-queue execution model continues to reduce synchronization costs as worker count increases.

\subsection{Conversational Retrieval Quality}
\label{subsec:retrieval_quality}

Finally, we evaluate whether memory-aware retrieval improves context-dependent follow-up queries. This experiment is not intended to show that HigressRAG is an inferior memory implementation; HigressRAG is evaluated in its standard stateless setting, where it receives only the follow-up query. OpRAG augments the query path with prior-turn context through \(Op_{memory}\).

In the large multicore conversational retrieval run, we evaluate 1195 valid follow-up cases generated from WikiText-103 source files. These queries are underspecified without prior context.
\begin{table}[t]
\centering
\caption{Conversational retrieval quality in the large multicore run.}
\label{tab:conv_quality}
\small
\begin{tabular}{lrrrr}
\toprule
System & Count & Top-1 & Hit@k & MRR \\
\midrule
OpRAG     & 1195 & 0.9560 & 1.0000 & 0.9730 \\
HigressRAG & 1195 & 0.0010 & 0.9940 & 0.0060 \\
\bottomrule
\end{tabular}
\end{table}
Table~\ref{tab:conv_quality} shows that prior-turn memory substantially improves Top-1 and MRR when follow-up queries depend on earlier context. This result evaluates the value of memory-aware retrieval, not a fully memory-enabled HigressRAG baseline. We also measure the overhead added by \(Op_{memory}\) by running the same 1195 queries with memory lookup disabled and enabled:
\[
\Delta T_{memory} =
T_{\text{with-memory}} - T_{\text{without-memory}}.
\]
The corresponding multicore timing summary shows that \(Op_{memory}\) adds only \textbf{0.03 ms/query} on average (\textbf{0.03 ms} p95(95th percentile) across repeated summary runs), or \textbf{2.0\%} of the retrieval-only query path. Thus, memory improves context-dependent retrieval quality while adding
only a small scheduled-operator cost to the critical path.

\subsection{CPU Strong and Weak Scaling}
\label{subsec:cpu_scaling}

The GPU experiments show that OpRAG improves end-to-end RAG execution with production-scale LLMs. We also evaluate CPU strong and weak scaling to isolate the behavior of the distributed operator runtime when the workload is dominated by preprocessing, embedding preparation, vector-index updates, and inter-stage coordination rather than GPU model execution. These experiments directly test whether the proposed operator runtime scales beyond the two-GPU setting.

The CPU scaling experiments run on a cluster where each node provides 40 physical CPU cores. The largest 1024-worker experiment uses 26 nodes, and all 1024 workers are physical workers rather than logical oversubscription. Strong scaling uses a fixed 100M-chunk corpus generated from \texttt{wikitext2\_train} and distributed across 4096 files. Weak scaling assigns 95K chunks to each worker, so the total workload grows with the number of workers. In both cases, the pipeline(Eq.~\ref{eq:four-pipeline}) executes the same four operators.

\begin{figure}[t]
\centering
\begin{tikzpicture}
\begin{groupplot}[
  group style = {
    group size = 2 by 2,
    horizontal sep = 0.9cm,
    vertical sep = 1.3cm
  },
  width=0.2\textheight,
  height=0.2\textheight,
  xmode=log,
  log basis x=2,
  xmin=128, xmax=1024,
  xtick={128,256,512,1024},
  xticklabels={128,256,512,1024},
  xticklabel style={font=\scriptsize, rotate=90, anchor=east},
  yticklabel style={font=\scriptsize},
  ylabel style={font=\scriptsize},
  xlabel style={font=\scriptsize},
  grid=both,
  legend cell align={center},
  legend style={
    font=\scriptsize,
    legend columns=2,
    at={(0.5,1.18)},
    anchor=south,
    draw=none,
    column sep=0.35cm
  }
]

\nextgroupplot[
  title={Load},
  ylabel={Time (s)},
  ymode=log
]

\addlegendentry{AsyncParallelOnly}
\addlegendentry{DaskScalableRAG}
\addlegendentry{HigressRAG}
\addlegendentry{OpRAG}

\addplot[blue, thick,mark=square*] coordinates {
  (128,4.300)(256,3.589)(512,5.362)(1024,6.500)
};

\addplot[orange,thick,mark=triangle*] coordinates {
  (128,8.668)(256,8.518)(512,8.790)(1024,7.252)
};

\addplot[purple,thick,mark=diamond*] coordinates {
  (128,1.427)(256,1.128)(512,0.507)(1024,0.337)
};

\addplot[green, thick,mark=otimes*] coordinates {
  (128,1.469)(256,0.876)(512,0.495)(1024,0.280)
};

\nextgroupplot[
  title={Transform},
  ymode=log
]

\addplot[blue, thick,mark=square*] coordinates {
  (128,26.248)(256,13.577)(512,6.709)(1024,3.936)
};

\addplot[orange,thick,mark=triangle*] coordinates {
  (128,12.429)(256,7.498)(512,4.868)(1024,3.017)
};

\addplot[purple,thick,mark=diamond*] coordinates {
  (128,7.723)(256,4.100)(512,2.715)(1024,2.093)
};

\addplot[green, thick,mark=otimes*] coordinates {
  (128,7.289)(256,5.267)(512,2.523)(1024,1.991)
};

\nextgroupplot[
  title={Embed},
  ylabel={Time (s)},
  xlabel={No of Workers},
  ymode=log
]

\addplot[blue, thick,mark=square*] coordinates {
  (128,89.521)(256,66.224)(512,24.667)(1024,13.528)
};

\addplot[orange,thick,mark=triangle*] coordinates {
  (128,42.996)(256,26.431)(512,13.556)(1024,6.090)
};

\addplot[purple,thick,mark=diamond*] coordinates {
  (128,20.646)(256,7.174)(512,4.526)(1024,2.717)
};

\addplot[green, thick,mark=otimes*] coordinates {
  (128,18.435)(256,8.523)(512,3.838)(1024,2.253)
};

\nextgroupplot[
  title={Upsert},
  xlabel={No of Workers},
  ymode=log
]

\addplot[blue, thick,mark=square*] coordinates {
  (128,10.388)(256,7.005)(512,2.431)(1024,1.647)
};

\addplot[orange,thick,mark=triangle*] coordinates {
  (128,152.708)(256,70.558)(512,38.585)(1024,19.707)
};

\addplot[purple,thick,mark=diamond*] coordinates {
  (128,5.670)(256,2.440)(512,2.402)(1024,1.678)
};

\addplot[green, thick,mark=otimes*] coordinates {
  (128,7.180)(256,3.104)(512,2.261)(1024,1.446)
};

\end{groupplot}
\end{tikzpicture}
\caption{Strong scaling behavior for Load, Transform, Embed, and Upsert.
The benchmark uses a fixed 100M-chunk corpus generated from
\texttt{wikitext2\_train} and distributed across 4096 files. Each node
has 40 physical CPU cores; the 1024-worker run uses 26 nodes and 1024
physical workers.} 
\label{fig:scaling_all}
\end{figure}
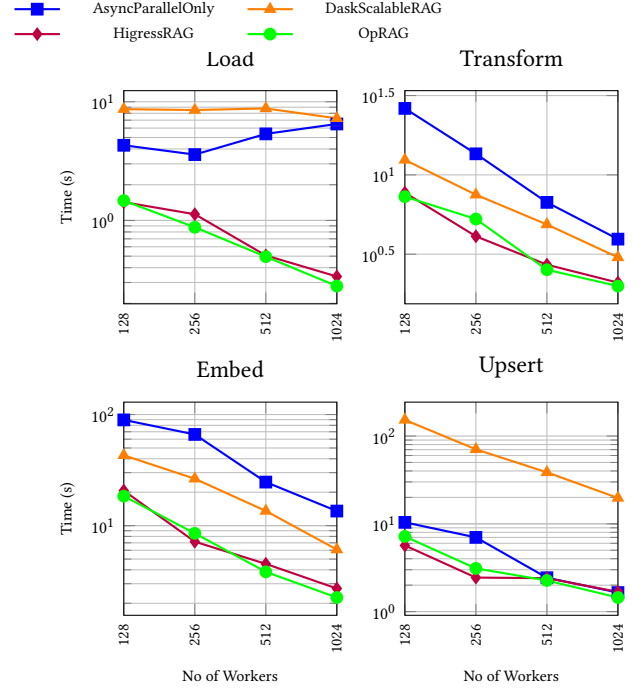

\begin{figure}[t]
\centering

\begin{tikzpicture}
\begin{groupplot}[
  group style = {
    group size = 2 by 1,
    horizontal sep = 0.9cm,
    vertical sep = 1.3cm
  },
  width=0.2\textheight,
  height=0.2\textheight,
  xmode=log,
  log basis x=2,
  xmin=128, xmax=1024,
  xtick={128,256,512,1024},
  xticklabels={128,256,512,1024},
  xticklabel style={font=\scriptsize, rotate=90, anchor=east},
  yticklabel style={font=\scriptsize},
  ylabel style={font=\scriptsize},
  xlabel style={font=\scriptsize},
  title style={font=\scriptsize, yshift=-2pt},
  grid=both,
  legend cell align={center},
  legend style={
    font=\scriptsize,
    legend columns=2,
    at={(0.5,1.18)},
    anchor=south,
    draw=none,
    column sep=0.35cm
  },
  xlabel={No of Workers}
]

\nextgroupplot[
  title={Strong Scaling Total},
  ylabel={Time (s)},
  ymode=log
]

\addlegendentry{AsyncParallelOnly}
\addlegendentry{DaskScalableRAG}
\addlegendentry{HigressRAG}
\addlegendentry{OpRAG}

\addplot[blue, thick,mark=square*] coordinates {
  (128,114.298)(256,83.266)(512,33.245)(1024,20.955)
};

\addplot[orange,thick,mark=triangle*] coordinates {
  (128,192.660)(256,95.486)(512,51.026)(1024,26.826)
};

\addplot[purple,thick,mark=diamond*] coordinates {
  (128,32.036)(256,14.576)(512,9.063)(1024,6.028)
};

\addplot[green, thick,mark=otimes*] coordinates {
  (128,30.944)(256,13.653)(512,7.514)(1024,4.505)
};

\nextgroupplot[
  title={Weak Scaling Total},
  ymode=log,
]

\addplot[blue, thick,mark=square*] coordinates {
  (128,17.564)(256,18.699)(512,26.276)(1024,27.107)
};

\addplot[orange,thick,mark=triangle*] coordinates {
  (128,24.214)(256,24.627)(512,27.562)(1024,39.506)
};

\addplot[purple,thick,mark=diamond*] coordinates {
  (128,5.228)(256,5.288)(512,4.833)(1024,6.015)
};

\addplot[green, thick,mark=otimes*] coordinates {
  (128,3.064)(256,3.150)(512,3.767)(1024,5.185)
};

\end{groupplot}
\end{tikzpicture}
\caption{Total runtime for CPU strong and weak scaling. Strong scaling
uses a fixed 100M-chunk workload, while weak scaling assigns 95K chunks
per worker. OpRAG remains the fastest configuration at 1024 physical
workers in both settings.} 
\label{fig:scaling_total_all}
\end{figure}
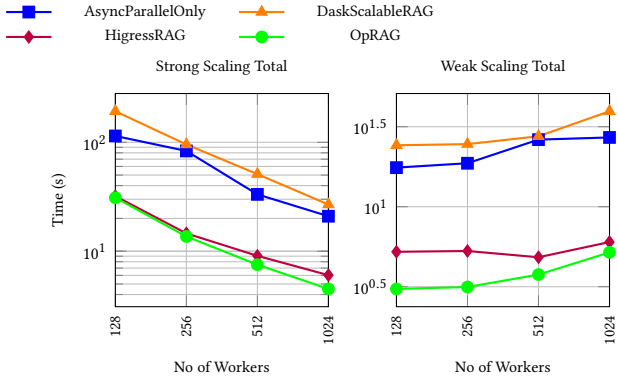

\textbf{Strong scaling:}
Figure~\ref{fig:scaling_all} reports stage-wise strong scaling for Load, Transform, Embed, and Upsert as the number of physical workers increases from 128 to 1024. The total runtime is shown separately in Fig.~\ref{fig:scaling_total_all}. OpRAG reduces total execution time from 30.944\,s at 128 workers to 4.505\,s at 1024 workers. This corresponds to a \(6.87\times\) reduction in total runtime as the worker count increases by \(8\times\). The closest baseline in strong scaling is HigressRAG. \textbf{At 1024 workers, HigressRAG requires 6.028\,s, while OpRAG requires 4.505\,s, which is $1.34\times$}.
The gap is larger against the other baselines: at 1024 workers, AsyncParallelOnly requires
20.955\,s and DaskScalableRAG requires 26.826\,s. The stage-wise trends in Fig.~\ref{fig:scaling_all} explain the total runtime improvement. In the Load stage, OpRAG decreases from 1.469\,s to 0.280\,s, while HigressRAG decreases from 1.427\,s to 0.337\,s. In Transform, OpRAG decreases from 7.289\,s to 1.991\,s. In Embed, the
dominant stage, OpRAG decreases from 18.435\,s to 2.253\,s, while HigressRAG decreases from 20.646\,s to 2.717\,s. In Upsert, OpRAG decreases from 7.180\,s to 1.446\,s, slightly faster than HigressRAG at 1024 workers. These stage-wise results support the design claim that OpRAG benefits from increasing physical parallelism because its operators are scheduled through bounded queues and persistent worker pools rather than through global stage barriers.

\begin{figure}[t]
\centering
\begin{tikzpicture}
\begin{groupplot}[
  group style = {
    group size = 2 by 2,
    horizontal sep = 0.9cm,
    vertical sep = 1.3cm
  },
  width=0.2\textheight,
  height=0.2\textheight,
  xmode=log,
  log basis x=2,
  xmin=128, xmax=1024,
  xtick={128,256,512,1024},
  xticklabels={128,256,512,1024},
  xticklabel style={font=\scriptsize, rotate=90, anchor=east},
  yticklabel style={font=\scriptsize},
  ylabel style={font=\scriptsize},
  xlabel style={font=\scriptsize},
  grid=both,
  legend cell align={center},
  legend style={
    font=\scriptsize,
    legend columns=2,
    at={(0.5,1.18)},
    anchor=south,
    draw=none,
    column sep=0.35cm
  }
]

\nextgroupplot[
  title={Load},
  ylabel={Time (s)},
  ymode=log
]

\addlegendentry{AsyncParallelOnly}
\addlegendentry{DaskScalableRAG}
\addlegendentry{HigressRAG}
\addlegendentry{OpRAG}

\addplot[blue, thick,mark=square*] coordinates {
  (128,3.903)(256,4.321)(512,7.148)(1024,9.272)
};

\addplot[orange,thick,mark=triangle*] coordinates {
  (128,8.964)(256,8.452)(512,9.200)(1024,10.006)
};

\addplot[purple,thick,mark=diamond*] coordinates {
  (128,1.921)(256,2.616)(512,2.251)(1024,3.142)
};

\addplot[green, thick,mark=otimes*] coordinates {
  (128,2.540)(256,2.882)(512,3.105)(1024,4.529)
};

\nextgroupplot[
  title={Transform},
  ymode=log
]

\addplot[blue, thick,mark=square*] coordinates {
  (128,4.201)(256,4.236)(512,3.289)(1024,4.653)
};

\addplot[orange,thick,mark=triangle*] coordinates {
  (128,2.548)(256,2.479)(512,3.035)(1024,3.020)
};

\addplot[purple,thick,mark=diamond*] coordinates {
  (128,0.878)(256,1.003)(512,1.718)(1024,1.869)
};

\addplot[green, thick,mark=otimes*] coordinates {
  (128,0.500)(256,0.617)(512,0.744)(1024,0.886)
};

\nextgroupplot[
  title={Embed},
  ylabel={Time (s)},
  xlabel={No of Workers},
  ymode=log
]

\addplot[blue, thick,mark=square*] coordinates {
  (128,13.339)(256,10.883)(512,19.886)(1024,19.568)
};

\addplot[orange,thick,mark=triangle*] coordinates {
  (128,7.047)(256,7.052)(512,6.079)(1024,9.051)
};

\addplot[purple,thick,mark=diamond*] coordinates {
  (128,2.773)(256,2.407)(512,2.073)(1024,2.749)
};

\addplot[green, thick,mark=otimes*] coordinates {
  (128,0.465)(256,0.415)(512,0.880)(1024,0.636)
};

\nextgroupplot[
  title={Upsert},
  xlabel={No of Workers},
  ymode=log
]

\addplot[blue, thick,mark=square*] coordinates {
  (128,0.835)(256,1.341)(512,1.858)(1024,1.637)
};

\addplot[orange,thick,mark=triangle*] coordinates {
  (128,15.014)(256, 16.398)(512,17.253)(1024,23.742)
};

\addplot[purple,thick,mark=diamond*] coordinates {
  (128,1.623)(256,1.355)(512,1.371)(1024,1.482)
};

\addplot[green, thick,mark=otimes*] coordinates {
  (128,0.066)(256,0.096)(512,0.109)(1024,0.082)
};

\end{groupplot}
\end{tikzpicture}
\caption{Weak scaling behavior for Load, Transform, Embed, and Upsert.
Each worker processes 95K chunks, so total workload size increases with
worker count. Each node has 40 physical CPU cores; the 1024-worker run
uses 26 nodes and 1024 physical workers.} 
\label{fig:scaling_weak_all}
\end{figure}
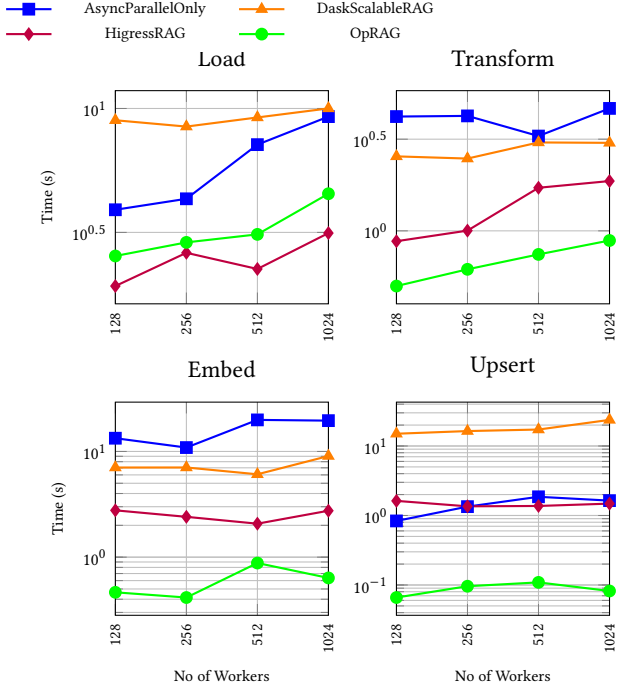

\textbf{Weak scaling:}
Figure~\ref{fig:scaling_weak_all} reports stage-wise weak scaling, and
Fig.~\ref{fig:scaling_total_all} reports total runtime. In weak scaling,
the workload grows with the worker count, so ideal behavior is not a
decrease in runtime but a shallow increase. OpRAG increases from
3.064\,s at 128 workers to 5.185\,s at 1024 workers, even though the
total number of chunks grows by \(8\times\). This moderate increase is
expected because metadata aggregation, global reductions, and upsert coordination grow with the number of partitions. OpRAG remains faster than all baselines across the weak-scaling range.
\textbf{At 1024 workers, the nearest baseline is HigressRAG at 6.015\,s, while OpRAG requires 5.185\,s which is $1.16\times$.}
Compared with AsyncParallelOnly and DaskScalableRAG, the gap is much larger: at 1024 workers, they require 27.107\,s and 39.506\,s, respectively. This shows that task parallelism alone is insufficient when growing workloads introduce more scheduler overhead, object
movement, and synchronization.

The stage-wise weak-scaling results in Fig.~\ref{fig:scaling_weak_all}
show why OpRAG remains stable. Transform grows only from 0.500\,s to
0.886\,s, and Embed remains below 1\,s at all scales, reaching 0.636\,s
at 1024 workers. Upsert remains nearly flat, increasing only from
0.066\,s to 0.082\,s. Load increases from 2.540\,s to 4.529\,s and
becomes the dominant contributor at large worker counts, indicating that
the remaining weak-scaling overhead is increasingly due to input
partition management rather than embedding or index-update execution.

Overall, the CPU scaling results support the same argument as the GPU
results. OpRAG's advantage is not caused by a model-specific
optimization. Instead, the runtime reduces the overhead term \(\Omega\) in equation \ref{eq:runtime_overtime}
by using operator-level scheduling, bounded queues, persistent workers, and explicit data movement. Strong scaling shows that OpRAG benefits from additional physical workers, while weak scaling shows that the runtime maintains stable execution as both data volume and worker count increase. Together, Figs.~\ref{fig:scaling_all},
\ref{fig:scaling_weak_all}, and \ref{fig:scaling_total_all} demonstrate that OpRAG's resource-deterministic execution model scales across hundreds to thousands of CPU workers.

\section{Related Work}
\label{sec:related}

OpRAG relates to distributed data systems, workflow runtimes, LLM programming frameworks, LLM serving systems, and memory-augmented RAG. Unlike prior systems that optimize one layer of the stack, OpRAG targets the orchestration layer around GPU-backed multi-stage RAG: embedding, retrieval, reasoning, memory, and upsert are treated as resource-aware operators with explicit scheduling and communication semantics.

\textbf{Distributed data systems and dataframe execution:}
Apache Spark~\cite{zaharia2016apache}, Apache Flink~\cite{carbone2015apache},
Dask~\cite{rocklin2015dask}, and Ray~\cite{moritz2018ray} provide
scalable batch, streaming, task-parallel, and AI-oriented execution
substrates. These systems are useful for large-scale preprocessing and
data movement, but their abstractions are not specialized for multi-stage RAG operators such as retrieval, memory lookup, context construction, and
vector-index upsert. Scalable dataframe systems such as
Modin~\cite{modin} and Cylon-based systems~\cite{widanage2020high,
abeykoon2022high,shan2022hybrid,alsaadi2025rhapsody,sarker2022incremental}
show that high-level data APIs can be lowered into communication-aware
execution plans. OpRAG builds on this direction, but applies the operator
view to GPU-backed RAG pipelines rather than dataframe analytics alone.

\textbf{Workflow runtimes and distributed execution control:}
Workflow systems such as RADICAL-Pilot and Parsl support task placement
and heterogeneous workflow execution on HPC platforms~\cite{merzky2022radical,
babuji2019parsl}. Google Pathways studies asynchronous distributed
dataflow for large ML programs~\cite{barham2022pathways}, and OneFlow
redesigns distributed execution for deep learning workloads
~\cite{yuan2021oneflow}. These systems separate program logic from
physical execution, but they treat RAG stages mostly as opaque tasks.
OpRAG adopts the same systems principle while exposing embedding,
retrieval, reasoning, memory, and upsert as explicit runtime operators
with known resource domains and communication patterns.

\textbf{LLM programming and agent frameworks:}
LangChain~\cite{langchain2023}, LangGraph~\cite{langgraph2024},
CrewAI~\cite{duan2024exploration,crewai_kickoff_async}, and
AutoGen~\cite{wu2024autogen} provide programming abstractions for tool
use, multi-agent coordination, and iterative reasoning. DSPy~\cite{dspy}
treats LM pipelines as optimizable programs. These systems improve
application-level programmability, but physical execution of embedding,
retrieval, memory access, and vector-index updates is typically delegated
to external libraries, vector stores, or callbacks. OpRAG is
complementary: it can serve as a deterministic execution substrate for
agentic or declarative RAG programs while preserving dynamic planning at
the framework layer.

\textbf{LLM serving runtimes:}
vLLM improves LLM serving throughput using PagedAttention and efficient KV-cache management~\cite{vllm, sarker2026aaflowplus}, while SGLang optimizes structured language-model programs through language/runtime co-design~\cite{sglang}. These systems optimize model execution, batching, decoding, and memory management inside the serving path. OpRAG does not replace these engines
or modify the decoding kernel. Instead, it optimizes the distributed dataflow around them: preprocessing, tokenization, embedding batches, retrieval, context construction, memory lookup, and index updates.

\textbf{RAG systems, vector retrieval, and memory:}
Classical RAG combines retrieval with generation for knowledge-intensive
tasks~\cite{lewis2020retrieval}. Practical RAG
systems rely on vector search and indexing backends such as
FAISS~\cite{johnson2019faiss}, ChromaDB, and Pinecone-style vector
stores~\cite{pinecone2023,hong2025benchmarking,hong2025context}.
Retrieval gateways such as HigressRAG optimize query-time routing,
hybrid retrieval, and semantic cache lookup~\cite{lin2025higress,
higress_repo,higress_cache}. Recent memory-augmented RAG systems,
including MemoRAG~\cite{qian2025memorag}, RAG-Tuned-LLM
~\cite{wei2025tuning}, HippoRAG~\cite{gutierrez2025rag}, and Cue
RAG~\cite{fu2025cue}, improve retrieval and memory policies for
long-context or multi-step reasoning. OpRAG is complementary to these
systems: rather than proposing only a new retrieval heuristic or memory
policy, it makes retrieval, memory, and upsert schedulable operators so
that latency, quality, and state-management tradeoffs can be measured in
one runtime.

\section{Discussion and Future Work}
\label{sec:discussion}

The evaluation shows that OpRAG improves GPU-backed multi-stage RAG by optimizing the orchestration layer around LLM execution rather than the LLM decoding kernel itself. Across both Llama3-8B and Mistral-7B, generation time remains similar across systems, while the largest performance differences appear in embedding, retrieval, context
construction, and vector-index update paths. This supports the central claim of the paper: in multi-stage RAG, end-to-end performance depends not only on model serving, but also on how surrounding dataflow stages are scheduled and coordinated.

\textbf{Why Operator-Level Execution Matters:} A consistent result across the GPU and CPU experiments is that simply adding parallelism is not sufficient. RayDataScalableRAG and DaskScalableRAG expose parallel execution, but their task scheduling, object movement, and stage coordination overheads remain visible in the end-to-end pipeline. HigressRAG is competitive for thin retrieval paths, especially when the query is self-contained and does not require memory or state updates. However, OpRAG performs better when the workload contains multiple interacting stages: embedding, retrieval, reasoning, memory, and upsert. This behavior follows the execution model in equation~\ref{eq:runtime_overtime}, 
where \(N\beta/P\) is useful work, \(N\alpha/(bP)\) is amortized per-batch overhead, and \(\Omega\) captures scheduler overhead, serialization, object-store traffic, synchronization, and idle time.
OpRAG improves performance primarily by reducing \(\Omega\). Persistent workers reduce repeated startup and dispatch overhead, bounded queues allow stages to overlap without unbounded buffering, and the Arrow/Cylon
data plane reduces serialization between preprocessing, embedding, retrieval, and indexing stages.

\textbf{GPU Results and CPU--GPU Overlap:} The GPU experiments show that OpRAG's gains persist with production-scale open models. In the pipeline and framework benchmarks, the dominant improvement comes from the embedding path rather than from generation. This is expected: OpRAG does not modify the decoding kernel. Instead, it
improves how inputs are prepared and delivered to the GPU. CPU tokenizer prefetching, length-aware batching, persistent GPU workers, and batched embedding reduce idle time and improve effective GPU utilization.

The results also show that upsert overhead can increase slightly in some GPU configurations because OpRAG performs more explicit batching and coordination before committing vectors. However, this overhead is small
relative to the embedding-stage savings, so the end-to-end pipeline still improves. This indicates that the runtime should be evaluated at the pipeline level rather than by optimizing each individual stage in
isolation.

\textbf{Retrieval Latency and Quality: } Full hybrid retrieval provides a strong quality reference because it combines dense and lexical evidence over the full candidate space. Both HigressRAG and OpRAG perform full-corpus hybrid retrieval with the same dense, lexical, score-fusion, and top-\(k\) semantics. The improvement comes from the physical execution plan rather than a change in retrieval behavior. OpRAG replaces document-wise Python BM25 scans with an exact inverted-postings implementation, uses vectorized
dense scoring, and applies partial top-\(k\) selection instead of full candidate sorting. These optimizations preserve the same retrieved results while reducing retrieval overhead. This shows that OpRAG's benefit comes from exposing retrieval and generation as runtime operators whose implementations can be optimized independently without changing the logical query semantics.

\textbf{CPU Scaling as Runtime Evidence:} The CPU scaling experiments complement the GPU evaluation. They isolate the distributed operator runtime when the workload is dominated by preprocessing, embedding preparation, vector-index update, and inter-stage coordination rather than GPU decoding. Strong scaling shows that OpRAG
benefits from additional physical workers, reducing total runtime as the worker count increases. Weak scaling shows that OpRAG maintains a shallow runtime increase as the total workload grows with the number of workers.

The scaling result supports the broader resource-deterministic execution argument. The same mechanisms that improve the GPU experiments---bounded queues, persistent workers, explicit operator scheduling, and reduced
serialization---also improve CPU scaling. Thus, OpRAG is not only a two-GPU optimization; it is a distributed runtime design that applies to both CPU-heavy and GPU-backed RAG pipelines.

\subsection{Limitations and Future Work}

OpRAG has several limitations. First, the current implementation uses FAISS as the primary vector backend. Although the operator abstraction can map to other stores, backend-specific behavior may affect upsert latency,
retrieval latency, and memory overhead. Second, the memory subsystem is evaluated through lookup, retrieval, and conversational quality metrics, but long-running memory growth and compaction are not yet fully characterized. Third, OpRAG currently focuses on data-parallel GPU execution with one model replica per GPU. It does not implement tensor parallelism, pipeline parallelism, or a new LLM decoding engine. Systems such as vLLM and SGLang remain complementary. Finally, OpRAG improves the orchestration layer, so its benefits are largest when preprocessing, embedding, retrieval, memory, or upsert contribute meaningfully to the critical path. If a workload is dominated almost entirely by long generation, the relative benefit will be smaller.

Our immediate future work is to practical extensions of the current system. First, while this paper demonstrates the macro-level end-to-end impact of the OpRAG execution model, a detailed micro-architectural breakdown isolating optimal queue depths, token-prefetch ratios, and serialization gains under highly variable adversarial workloads is left for future work. Second, we plan to measure memory growth under longer multi-turn workloads and report how retention policy, summary count, and embedding dimension affect memory footprint. 
Finally, we plan to add more repeated runs and confidence intervals for the largest GPU and CPU experiments to better quantify variance. These extensions do not change the scope of OpRAG. The goal remains to provide a resource-deterministic runtime for the orchestration layer of GPU-backed multi-stage RAG, complementing rather than replacing existing LLM serving systems.

\section{Conclusion}
\label{sec:conclusion}

This paper presented OpRAG, a resource-deterministic runtime for GPU-backed multi-stage RAG workflows. OpRAG treats embedding, retrieval, reasoning, memory, and upsert as first-class operators and executes
agent-selected segments through deterministic, communication-aware runtime plans. By combining persistent workers, bounded queues, CPU--GPU overlap, and a serialization-free Arrow/Cylon data plane, OpRAG optimizes the orchestration layer around LLM execution without modifying the decoding kernel.

Our evaluation with Llama3-8B and Mistral-7B on A100 GPUs shows that OpRAG improves end-to-end pipeline time over the nearest competitor by 16.16\% and 15.66\%, respectively, and achieves more than 20\% improvement over RayScalableRAG. Framework-level GPU comparisons, Higress-style query serving, and CPU strong/weak scaling further show that the same operator-runtime design reduces non-model overhead across embedding, retrieval, memory, and index-update paths. These results suggest that scalable multi-stage RAG systems should expose retrieval, memory, and indexing as schedulable systems operators rather than hidden framework callbacks.

\bibliographystyle{ACM-Reference-Format}
\bibliography{OpRAG}

\end{document}